\documentclass[aps,prd,reprint,groupedaddress,showpacs]{revtex4-1}

\usepackage{graphicx}
\usepackage{amsmath}

\begin{document}

\title{``Big Bang'' as a first-order phase transition in the early Universe}


\author{E.A. Pashitskii}
\email[pashitsk@iop.kiev.ua]{}
\affiliation{Institute of Physics, NAS of Ukraine, 46 Nauki Avenue, Kiev, 03028, Ukraine}


\date{May 16, 2014}

\begin{abstract}
It is argued that the ``Big Bang'' initiating the creation of our Universe may be a consequence of a first-order phase transition induced by interaction of a fundamental non-linear scalar field with gravitational field. The Lagrangian describing the scalar field $\phi $ characterized by ``imaginary mass'' and nonlinearity of $\phi ^{4} $ type, existing in the space-time with non-zero scalar curvature $R$, is proposed to be augmented with an additional linear term ${\beta R\phi \mathord{\left/ {\vphantom {\beta R\phi  \sqrt{\kappa } }} \right. \kern-\nulldelimiterspace} \sqrt{\kappa } } $, along with the standard term $-\xi R\left|\phi \right|^{2} $ quadratic in $\phi $, where $\xi $ and $\beta $ are dimensionless constants and $\kappa $ is the Einstein's gravitational constant. The term linear in $\phi $, playing the role of an ``external field'', leads to a cubic equation in $\phi $ for the extrema of the potential energy of the scalar field and ensures the possibility of a first-order phase transition driven by the parameter proportional to $R$. It is assumed that the early Universe is filled with non-linear scalar field in the ground state and cold matter, neutral with respect to all charges, satisfying the equation of state $p=\nu \varepsilon $. It is shown that given the condition $\nu >{1\mathord{\left/ {\vphantom {1 3}} \right. \kern-\nulldelimiterspace} 3} $ the scalar curvature $R=\kappa \left(3\nu -1\right)\varepsilon -4\Lambda $ (where $\Lambda $ is the cosmological constant) decreases with diminishing of the energy density of matter during the Universe's expansion and reaches certain critical value $R_{c} <0$ when the first-order phase transition occurs. Using parameters characterizing the Higgs field, the rapid ``roll-down'' of the system into the potential minimum is shown to take place in a time span of about $10^{-31}$~s. During this time the latent heat of the transition is released increasing the temperature of the Universe to the Planck value $T_{P} \approx 10^{32} $~K, which may be seen as the ``Big Bang'' producing $4 \cdot 10^{30}$~GW of power.
\end{abstract}


\maketitle

\section{Introduction}
In the contemporary cosmological models the concept of ``Big Bang'' is actually supplunted by a notion of a rapid chaotic inflationary expansion of the early Universe due to the anomalously large (of the Planck order of magnitude) vacuum energy density which is equal to the potential energy of a fundamental scalar field \cite{Linde} $V(\varphi ) \sim M_P^4$ ($ M_P$ is Planck mass). There is though a problem with the mechanism of the Universe's heating to high (Planck) temperature $ T_P \sim M_P \sim {10^{19}}$~GeV due to the sacalar field's fluctuations, as the energy of these fluctuations after inflation do not exceed $10^{12}$~GeV.

An alternative scenario of the origination of our Universe in the process of a first-order phase transition driven by interaction between nonlinear scalar field and gravitational field is suggested in this paper. It is proposed to supplement the Lagrangian of a fundamental scalar field $\phi $, having ``imaginary mass'' and nonlinearity of $\phi ^{4} $ type, with an ``external field'' term ${\beta R\phi \mathord{\left/ {\vphantom {-\beta R\phi  \sqrt{\kappa } }} \right. \kern-\nulldelimiterspace} \sqrt{\kappa } } $ linear in $\phi $ and scalar curvature $R$, along with the usual term $\propto R\phi ^{2} $ quadratic in $\phi $ (see \cite{Krive,Bezrukov}). The linear term contains, besides the dimensionless constant $\beta $, a dimensional multiplier ${1\mathord{\left/ {\vphantom {1 \sqrt{\kappa } }} \right. \kern-\nulldelimiterspace} \sqrt{\kappa } } $, where $\kappa $ is the Einstein's gravitational constant with dimensionality coinciding with that of $\phi ^{-2} $. Such a term leads to a cubic equation in $\phi $ for the extrema of the potential energy $U(\phi )$ and ensures the possibility of the first-order phase transition \cite{Landau}. Three real roots of the cubic equation, i.e. three real values of the scalar field $\phi $, corresponding to two minima and one maximum of the potential energy exist for a certain range of the values of $R$. We consider a spatially closed Universe filled with nonlinear scalar field $\phi $ in the ground state and with cold matter described by the equation of state $p=\nu \varepsilon $, where $p$ and $\varepsilon $ are the pressure and energy density of matter and $\nu >{1\mathord{\left/ {\vphantom {1 3}} \right. \kern-\nulldelimiterspace} 3} $. In this case the scalar curvature, given a nonzero value of cosmological constant $\Lambda $, equals $R=\kappa \left[\left(3\nu -1\right)\varepsilon -4\lambda \right]$, where $\lambda ={\Lambda \mathord{\left/ {\vphantom {\Lambda  \kappa }} \right. \kern-\nulldelimiterspace} \kappa } $ is the energy density of vacuum.

It is shown below that for the initial condition $\left(3\nu -1\right)\varepsilon _{0} >4\lambda $ the positive scalar curvature $R>0$ in time becomes negative as $\varepsilon $ diminishes in the course of the Universe's expansion. At a certain moment $t_{c} $, when the radius of the Universe $a\left(t\right)$ and scalar curvature $R\left(t\right)$ reach critical values $a_{c} =a\left(t_{c} \right)$ and $R_{c} =R\left(t_{c} \right)<0$, the first-order phase transition occurs, accompanied by the release of the latent heat of transition which equals the change in the potential energy $\Delta U$ of the scalar field. With the Higgs field \cite{Higgs1,*Higgs2} taken as an example of the nonlinear scalar field $\phi $ it is possible to evaluate the model parameters as well as the characteristic times of the inflationary expansion $t_{\inf } \sim 10^{-21} $~s and phase transition $t_{1} \sim 10^{-31} $~s. Thus, on the backdrop of a rather slow inflation of the early cold Universe an almost instantaneous first-order phase transition occurs with rapid heating of matter in the Universe to ultrahigh temperature of the order of the Planck one $T_{P} \sim 10^{32} $~K, i.e. the ``Big Bang'' with power of about $10^{30} $~GW, .

\vspace{-3mm}

\section{Modified Lagrangian of the real nonlinear scalar field in curved space-time and the first-order phase transition}

Let us consider a modified Lagrangian of some real non-linear scalar field $\phi $ with ``imaginary mass'' $\mu $ and nonlinearity of $\phi ^{4} $ type in a curved 4-space with metric tensor $g^{\mu \nu } $ and non-zero scalar curvature $R\ne 0$:

\begin{equation} \label{Eq1} L=g^{\mu \nu } (\partial _{\mu } \phi )(\partial _{\nu } \phi )+\frac{1}{2} \mu ^{2} \phi ^{2} -\frac{1}{4} g^{2} \phi ^{4} -\xi R\phi ^{2} +\frac{\beta R}{\sqrt{\kappa } } \phi .  \end{equation}

The Lagrangian (\ref{Eq1}) differs from a standard one, containing a term quadratic in $\phi $ and linear in $R$ with some dimensionless constant $\xi $ (see \cite{Krive,Bezrukov}), by and additional term linear in both $\phi $ and $R$, acting as an ``external field''. The physical dimension of $\phi ^{2} $ coincides with the dimension of the reciprocal Einstein's constant $\kappa ^{-1} ={c^{4} \mathord{\left/ {\vphantom {c^{4}  8\pi G_{N} }} \right. \kern-\nulldelimiterspace} 8\pi G_{N} } $ (where $G_{N} $ is the Newton's gravitational constant and $c$ is the velocity of light), so the last term in (\ref{Eq1}) has a dimensional multiplier ${1\mathord{\left/ {\vphantom {1 \sqrt{\kappa } }} \right. \kern-\nulldelimiterspace} \sqrt{\kappa } } $ along with a dimensionless constant $\beta $.

Expression (\ref{Eq1}) implies that the potential energy of the real field $\phi $ is:
\begin{equation} \label{Eq2} U(\phi ,R)=-\frac{(\mu ^{2} -\xi R)}{2} \phi ^{2} +\frac{g^{2} }{4} \phi ^{4} -\frac{\beta R}{\sqrt{\kappa } } \phi +U_{0} ,  \end{equation}
where constant $U_{0} $ is chosen so that the minimal value of $U$ equals zero in the ground state.

The extrema of potential (\ref{Eq2}) are given by cubic equation with respect to $\phi $:
\begin{equation} \label{Eq3} \frac{\partial U}{\partial \phi } =g^{2} \phi ^{3} -(\mu ^{2} -\xi R)\phi -\frac{\beta R}{\sqrt{\kappa } } =0.  \end{equation}

In a particular case of flat space-time ($R=0$) equation (\ref{Eq3}) gives the standard expression for the vacuum average of the scalar field's amplitude $\phi _{0} ={\mu \mathord{\left/ {\vphantom {\mu  g}} \right. \kern-\nulldelimiterspace} g} $, while from (\ref{Eq2}) and with the choice of the constant $U_{0} ={\mu ^{2} \phi _{0}^{2} \mathord{\left/ {\vphantom {\mu ^{2} \phi _{0}^{2}  4}} \right. \kern-\nulldelimiterspace} 4} $ we obtain zero equilibrium energy density of the scalar field at the points $\phi =\pm \phi _{0} $ (see \cite{Krive}).

For the farther analysis it is convenient to rewrite equations (\ref{Eq2}) and (\ref{Eq3}) introducing dimensionless variables $x={\phi \mathord{\left/ {\vphantom {\phi  \phi _{0} }} \right. \kern-\nulldelimiterspace} \phi _{0} } $ and $V={U\mathord{\left/ {\vphantom {U \mu ^{2} \phi _{0}^{2} }} \right. \kern-\nulldelimiterspace} \mu ^{2} \phi _{0}^{2} } $:
\begin{equation} \label{Eq4} V(x,h)=\frac{1}{4} x^{4} -\frac{1}{2} (1-\alpha h)x^{2} - hx;  \end{equation}
\begin{equation} \label{Eq5} x^{3} -(1-\alpha h)x - h=0,  \end{equation}
where
\begin{equation} \label{Eq6)} \alpha =\phi _{0} \sqrt{\kappa } \cdot \xi /\beta ; \quad  h=\beta R/\mu ^{2} \phi _{0} \sqrt{\kappa } =\xi R/\alpha \mu ^{2} .  \end{equation}

\begin{figure}
\includegraphics[width=\columnwidth]{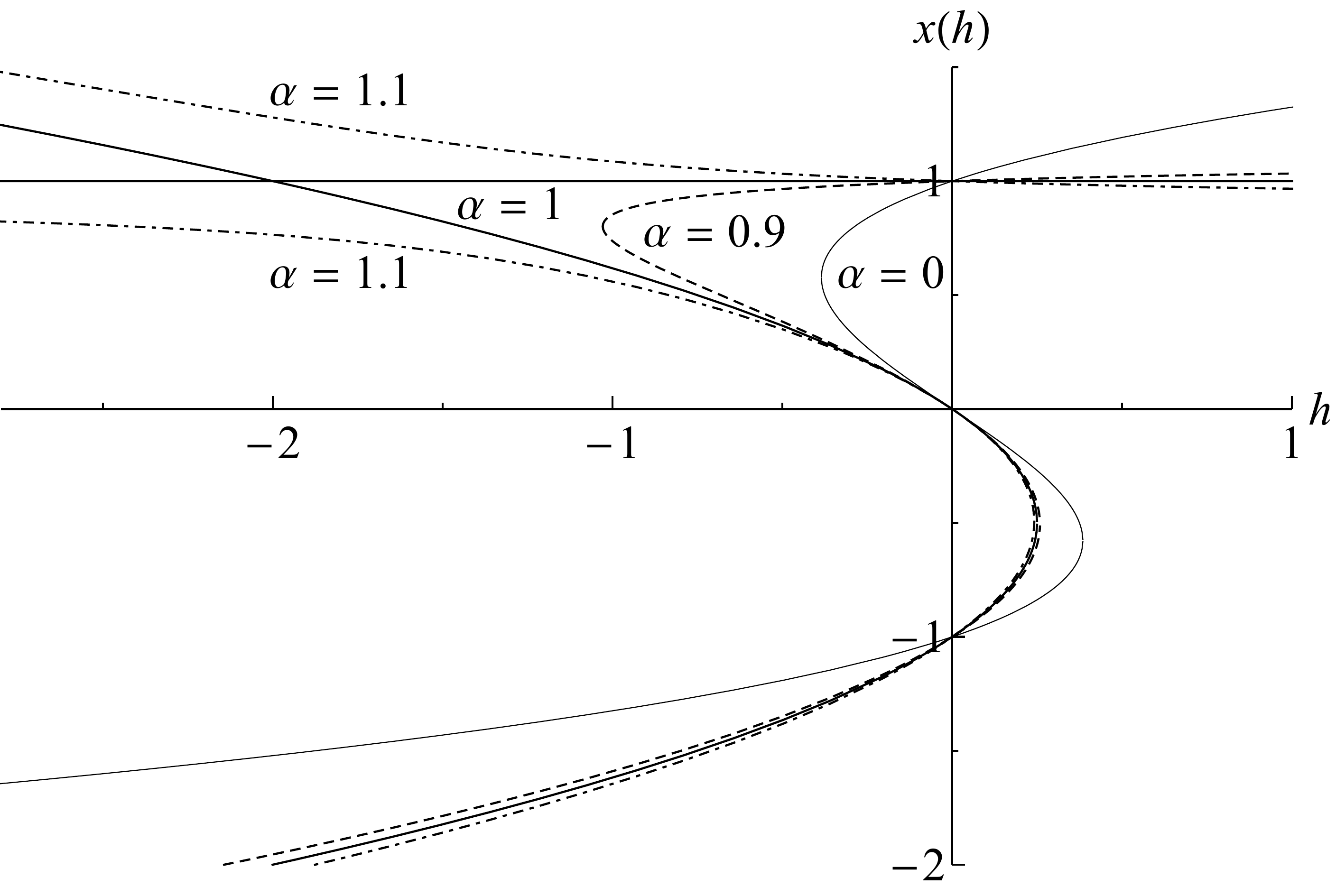}
\caption{\label{Fig1} The $h$ dependencies of the real roots of the cubic equation (\ref{Eq5}) for several values of the dimensionless parameter $\alpha $. The value $\alpha =1$ is critical for the existence of the first-order phase transition: for $\alpha >1$ there are always three real roots corresponding to two minima and one maximum, so that the first-order phase transition is impossible (at zero temperature system can not leave one of the minima).}
\end{figure}


Fig.~\ref{Fig1} shows the $h$ dependencies of the real roots of cubic equation (\ref{Eq5}) for several values of the dimensionless parameter $\alpha $. For $0\le \alpha \le 1$ there are three real roots in the region $-h_{1} \left(\alpha \right)<h<h_{2} \left(\alpha \right)$, corresponding to two (meta)stable states at the minima of the potential (\ref{Eq4}), and an unstable state at the maximum.

According to (\ref{Eq4}) and (\ref{Eq5}) the first-order phase transition is possible in this range of $\alpha$. For $\alpha =1$, when $h_{1} \left(1\right)=2$ and $h_{2} \left(1\right)=0.5$, one of the roots is constant $x=1$ for all values of $h$, which corresponds to $\phi =\phi _{0} $, while for $\alpha <1$, including the case of $\alpha =0$ with $h_{1} \left(0\right)=h_{2} \left(0\right)\approx 0.385$, the values of all roots depends on $h$.

For $\alpha >1$ there are three real roots in the whole region $h<h_{2} \left(\alpha \right)$. It means that potential (\ref{Eq4}) always has two minima divided by a potential barrier (maximum), so that the classical system always resides in one of the minima and the first-order phase transition is impossible.

\begin{figure}[b]
\includegraphics[width=\columnwidth]{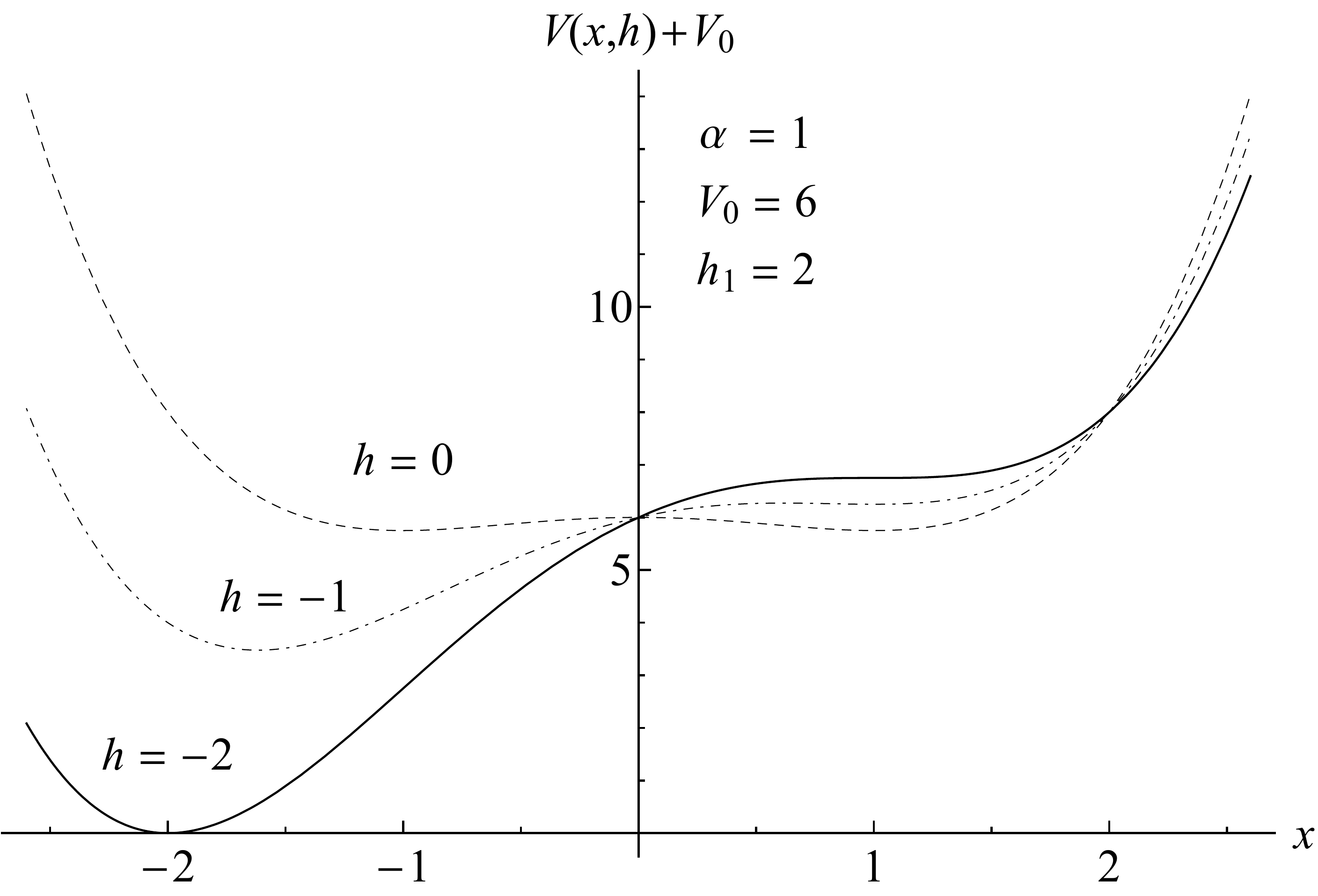}
\caption{\label{Fig2} The $x$ dependencies of the potential $V(x,h)$ for $\alpha =1$ and for several values of the dimensionless parameter $h$: $h=0$ (dotted curve), $h=-1$ (dashed curve) and $h=-h_{1} =-2$ (solid curve).}
\end{figure}

Thereby the value $\alpha =1$, which implies $\phi _{0} \sqrt{\kappa } ={\beta \mathord{\left/ {\vphantom {\beta  \xi }} \right. \kern-\nulldelimiterspace} \xi } $, is the extreme (maximal) value of the parameter $\alpha $ still allowing the first-order phase transition. The value of the order parameter for the nonlinear Higgs field $\phi _{0} ={\mu \mathord{\left/ {\vphantom {\mu  g}} \right. \kern-\nulldelimiterspace} g} \approx 247$~GeV \cite{Weinberg} (in the units $\hbar =c=1$) leads to the estimate of the dimensionless quantity $\phi _{0} \sqrt{\kappa } \approx \sqrt{{G_{N} \mathord{\left/ {\vphantom {G_{N}  G_{F} }} \right. \kern-\nulldelimiterspace} G_{F} } } \approx 10^{-16} $ (where $G_{F} $ is the Fermi constant in the theory of weak interaction), so the parameter $\alpha $ may take value of $\alpha \approx 1$ only for ${\xi \mathord{\left/ {\vphantom {\xi  \beta }} \right. \kern-\nulldelimiterspace} \beta } \sim 10^{16} $.

In Fig.~\ref{Fig2} the $x$ dependencies of the potential $V\left(x\right)$ are shown for $\alpha =1$ and for several values of the dimensionless parameter $h$. For $h=-h_{1} $ the right minimum of the potential $V\left(x\right)$ and maximum at $x=0$ disappear, leaving only one minimum at $x=-2$.

\begin{figure}
\includegraphics[width=\columnwidth]{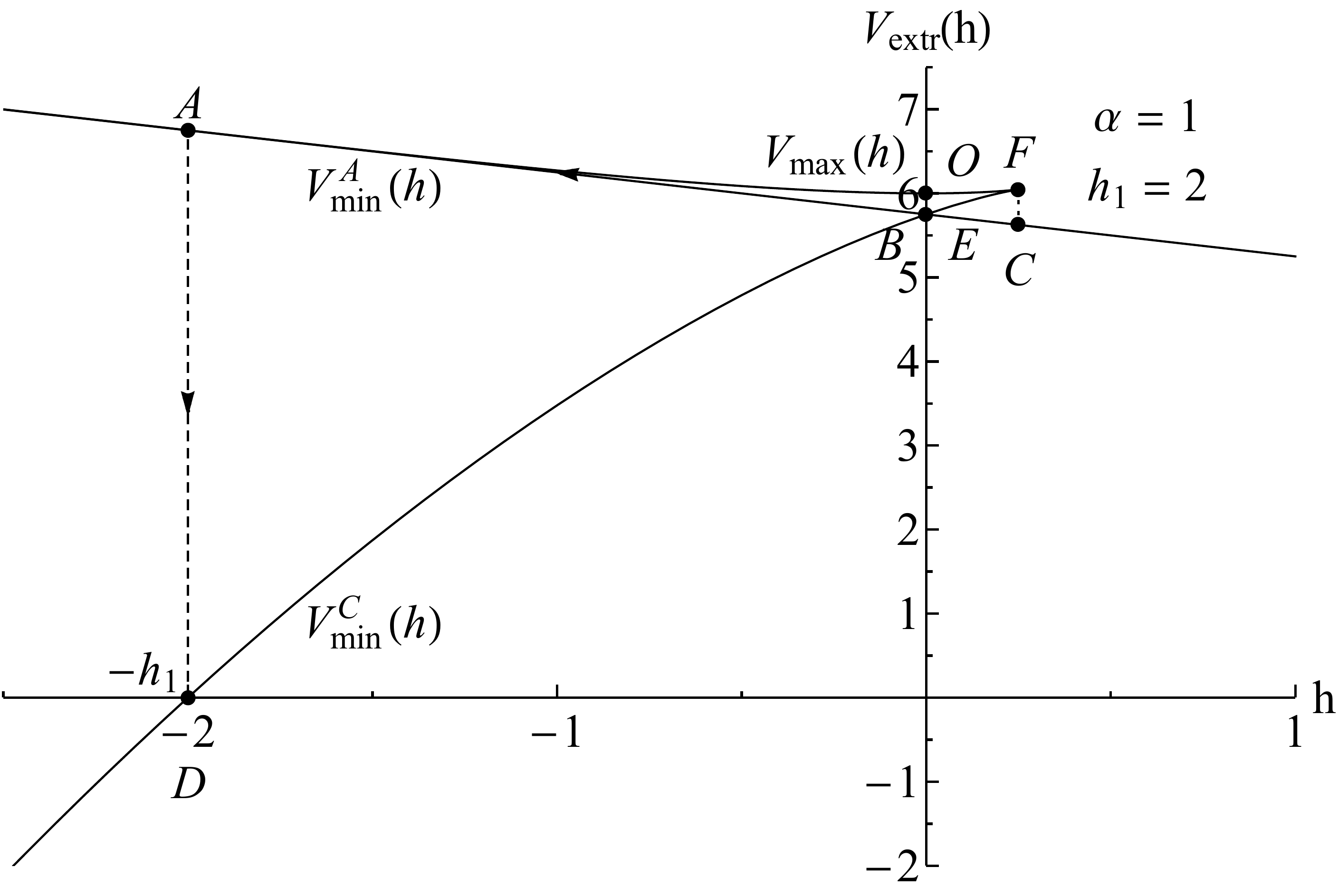} \caption{\label{Fig3} The $h$ dependencies of the extremal values of the potential $V_{extr} (h)$ for $\alpha =1$, indicating the first-order phase transition from point $A$ to point $D$ with the release of latent energy $\Delta V$.} \end{figure}

Fig.~\ref{Fig3} represents the $h$ dependencies of the extremal values of the potential (\ref{Eq4}) for $\alpha =1$. The branches $ABC$ and $DEF$ correspond to metastable states in the two minima of the potential, while the branch $AOF$ depicts absolutely unstable states at the maximum of the potential. At $h=0$ the depths of the two minima are equal (points $B$ and $E$ coincide). It will be shown below that under certain conditions the early Universe in the course of its evolution may travel along the branch $CBA$ up to the point of the first-order phase transition $A$ at $h=-h_{1} =-2$, where it jumps to the point $D$ with the drop $\Delta V=6.75$ in the dimensionless potential. This corresponds to the change $\Delta U=6.75\mu ^{2} \phi _{0}^{2} $ of the density of potential energy of the scalar field.

The total energy released in the first-order phase transition depends on the critical values of the radius $a_{c} $ and volume $\upsilon _{c} =2\pi ^{2} a_{c}^{3} $ of spatially closed early Universe at the point of transition. Assuming that the whole released energy is spent on heating of the early Universe to the temperature of about the Planck temperature $T_{P} \approx 1.4\cdot 10^{32} K$, the relation $k_{B} T_{P} \approx \Delta U\cdot \upsilon _{A} $ provides estimates of the critical radius and volume of the Universe at the transition point if the parameters of the scalar field are known.

In particular, for the Higgs field with $\mu ={m_{B} \mathord{\left/ {\vphantom {m_{B}  \sqrt{2} }} \right. \kern-\nulldelimiterspace} \sqrt{2} } $, where $m_{B} \approx 125$~GeV is the mass of the Higgs boson observed in the Large Hadron Collider experiments \cite{ATLAS,CMS}, for the phase transition $AD$ (Fig.~\ref{Fig3}) we have $\Delta U\approx 3.24\cdot 10^{9} {\rm \; GeV}^{4} $ (in the units $\hbar =c=1$), or $\Delta U\approx 6.79\cdot 10^{47} {\rm \; }{{\rm erg}\mathord{\left/ {\vphantom {{\rm erg} {\rm cm}^{{\rm 3}} }} \right. \kern-\nulldelimiterspace} {\rm cm}^{{\rm 3}} } $ (in CGS units). The condition $\Delta U\cdot \upsilon _{c} \approx k_{B} T_{P} \approx 2\cdot 10^{16} $~erg gives $\upsilon _{c} \approx 3\cdot 10^{-32} {\rm \; cm}^{3} $ and $a_{c} \approx 10^{-11} {\rm \; cm}$.

\section{Evolution of the early Universe towards the first-order phase transition}
It is assumed hereafter that the first-order phase transition in the early Universe occurs when the time-dependent dimensionless parameter $h\left(t\right)$, proportional to the scalar curvature $R\left(t\right)$, at some moment of time $t=t_{c} $ reaches a negative critical value $h_{c} \equiv h\left(t_{c} \right)=-h_{1} <0$ ($h_{1} =2$ for $\alpha =1$). Let us show that under certain conditions the Universe in the course of its evolution may reach this point of phase transition moving with time along the phase trajectory $CBA$ (see Fig.~\ref{Fig3}) from some initial point to the point $A$.

In order to describe the evolution of the isotropic and homogeneous spatially closed Universe filled with cold matter, governed by the equation of state $p=\nu \varepsilon $ with $\nu >{1\mathord{\left/ {\vphantom {1 3}} \right. \kern-\nulldelimiterspace} 3} $, let us use the classical equations of the general theory of relativity with noon-zero cosmological constant (see \cite{Zeldovich}):
\begin{equation} \label{Eq7} \dot{a}^{2} +c^{2} =\frac{\kappa c^{2} }{3} (\varepsilon +\lambda )a^{2} ; \quad  \ddot{a}=-\frac{\kappa c^{2} }{6} [(3\nu +1)\varepsilon -2\lambda ]a, \end{equation}
where $a$ is a time-dependent spatial scale (radius) of the Universe.

Equations (\ref{Eq7}) give the energy conservation law (for $\lambda =const$):
\begin{equation} \label{Eq8} 3\frac{\dot{a}}{a} +\frac{\dot{\varepsilon }}{(1+\nu )\varepsilon } =0,  \end{equation}
as well as the expression for scalar curvature:
\begin{equation} \label{Eq9} R=-\frac{6}{c^{2} a^{2} } (a\ddot{a}+\dot{a}^{2} +c^{2} )=\kappa [(3\nu -1)\varepsilon -4\lambda ].  \end{equation}

According to the energy conservation law (\ref{Eq8}) the growth of $a\left(t\right)$ in the process of expansion of the Universe results in the power-law decreasing of the energy density of matter:
\begin{equation} \label{Eq10} \varepsilon (t)=\varepsilon _{0} \cdot [a_{0} /a(t)]^{3(1+\nu )} ,  \end{equation}
where $\varepsilon _{0}$ and $a_{0}$ are initial values.

Equations (\ref{Eq7}) may be rewritten then as
%
\begin{equation}
\begin{split}
\label{Eq11} \dot{a}^{2} (t) &=c^{2} \left\{\frac{\Lambda }{3} \left[1+\chi \left(\frac{a_{0} }{a(t)} \right)^{3(1+\nu )} \right]a^{2} (t)-1\right\}; \\ \quad  \ddot{a}(t) &=\frac{c^{2} \Lambda }{3} \left[1-\frac{3}{2} \chi \left(\frac{a_{0} }{a(t)} \right)^{3(1+\nu )} \right]a(t),
\end{split}
\end{equation}
%
where $\chi ={\varepsilon _{0} \mathord{\left/ {\vphantom {\varepsilon _{0}  \lambda }} \right. \kern-\nulldelimiterspace} \lambda } $. The first of the equations (\ref{Eq11}) implies that the expansion is possible only if the condition ${\Lambda a_{0}^{2} \left(1+\chi \right)\mathord{\left/ {\vphantom {\Lambda a_{0}^{2} \left(1+\chi \right) 3}} \right. \kern-\nulldelimiterspace} 3} >1$ is satisfied. After some moment $t^{*}, $ when $a^{*} =a\left(t^{*} \right)\gg a_{0} $ and $\chi \cdot \left({a_{0} \mathord{\left/ {\vphantom {a_{0}  a^{*} }} \right. \kern-\nulldelimiterspace} a^{*} } \right)^{3\left(1+\nu \right)} \ll 1$, the solution of (\ref{Eq11}) settles on the inflationary (exponential) regime of the Universe's expansion:
\begin{equation} \label{Eq12} a(t)\approx a^{*} \cdot \exp \left\{c(t-t^{*} )\sqrt{\Lambda /3} \right\}.  \end{equation}

Thus according to (\ref{Eq10}) the energy density of matter $\varepsilon \left(t\right)$ decreases rapidly with time, so that scalar curvature (\ref{Eq9}) and dimensionless parameter $h$ in (\ref{Eq5}) become negative with the system (the early Universe) approaching the point of first-order phase transition.

The duration of the first-order phase transition, i.e the time of the system's ``roll-down'' into the point of the absolute minimum (see Fig.~\ref{Fig2}), may be estimated using the expression for the scalar field's evolution:
\begin{equation} \label{Eq13)} 3H\dot{\phi }=-c^{2} \partial U(\phi )/\partial \phi ,  \end{equation}
where $H={\dot{a}\mathord{\left/ {\vphantom {\dot{a} a}} \right. \kern-\nulldelimiterspace} a} $ is the Hubble constant, while the derivative of the potential is given by (\ref{Eq3}). Thus during the inflationary stage, when $H=c\sqrt{{\Lambda \mathord{\left/ {\vphantom {\Lambda  3}} \right. \kern-\nulldelimiterspace} 3} } $, taking the derivative in the point $\phi =0$ and assuming $\dot{\phi }\approx {\phi _{0} \mathord{\left/ {\vphantom {\phi _{0}  t}} \right. \kern-\nulldelimiterspace} t} $, we arrive at the following estimate for the duration of the first-order phase transition:
\begin{equation} \label{Eq14} t_{1} \approx \sqrt{3\Lambda } /2c\mu ^{2} .  \end{equation}

On the other hand, according to (\ref{Eq12}) the characteristic time of inflation is
\begin{equation} \label{Eq15} t_{\inf } =c^{-1} \sqrt{3/\Lambda } .  \end{equation}

We can see that both times depend on the value of $\Lambda $ in the early Universe. To estimate $\Lambda $ we may assume that, on the one hand, the critical value of the scalar curvature $R_{c} $ at the point of transition is close to $-4\Lambda $ due to the smallness of the energy density of matter $\varepsilon _{c} \equiv \varepsilon \left(t_{c} \right)$ while, on the other hand, its absolute value by the order of magnitude is $\left|R_{c} \right|\approx {1\mathord{\left/ {\vphantom {1 a_{c}^{2} }} \right. \kern-\nulldelimiterspace} a_{c}^{2} } $. This gives $\Lambda \approx {1\mathord{\left/ {\vphantom {1 4}} \right. \kern-\nulldelimiterspace} 4} a_{c}^{2} \approx 2.5\cdot 10^{21} {\rm \; cm}^{2} $, so that $\lambda ={\Lambda \mathord{\left/ {\vphantom {\Lambda  \kappa }} \right. \kern-\nulldelimiterspace} \kappa } \approx 10^{69} {\rm \; }{{\rm erg}\mathord{\left/ {\vphantom {{\rm erg} {\rm cm}^{3} }} \right. \kern-\nulldelimiterspace} {\rm cm}^{3} } $.

As the result, from (\ref{Eq14}) and (\ref{Eq15}) we have $t_{1} \approx 1.4\cdot 10^{-31} $~s and $t_{\inf } \approx 10^{-21} $~s, which means that the first-order phase transition happens almost instantly on the background of the much slower inflation of the early cold Universe.

In  the same time, the relation $\left|R_{c} \right|={2\mu ^{2} \mathord{\left/ {\vphantom {2\mu ^{2}  \xi }} \right. \kern-\nulldelimiterspace} \xi } $ provides estimates for the interaction constants of the Higgs field with gravitational field: $\xi \approx 2\mu ^{2} a_{c}^{2} \approx 4\cdot 10^{9} $ and $\beta \approx 4\cdot 10^{-7} $.

To summarise, the first-order phase transition with the release of the latent heat $\Delta E=\Delta U\cdot \upsilon _{c} $ comparable to the Planck energy $k_{B} T_{P} =M_{P} c^{2} \approx 1.2\cdot 10^{19} $~GeV during the time of about $10^{-31} $~s plays the role of ``Big Bang'' with the power of about $W\approx 4\cdot 10^{30} $~GW, which is $10^{13} $ times the power of the Sun's radiation.

\vspace{5mm}

\section{Conclusion}
A new scenario is proposed for the inception of our Universe in the course of the first-order phase transition, initiated by interaction of non-linear scalar field $\phi $ with gravitational field in the presence of cold neutral matter governed by the equation of state $p=\nu \varepsilon $ with $\nu >{1\mathord{\left/ {\vphantom {1 3}} \right. \kern-\nulldelimiterspace} 3} $. The origins of this matter, as well as the origins of the scalar field, are not discussed herein. It is assumed that the Lagrangian of the scalar field in the curved 4-space with non-zero scalar curvature contains, along with the standard term $-\xi R\phi ^{2} $ quadratic in $\phi $, an ``external field'' term ${\beta R\phi \mathord{\left/ {\vphantom {-\beta R\phi  \sqrt{\kappa } }} \right. \kern-\nulldelimiterspace} \sqrt{\kappa } } $ linear in $\phi $, which under certain conditions ensures first-order phase transition driven by the parameter proportional to $R$. The scalar curvature, which equals in this case to $R=\kappa \left(3\nu -1\right)\varepsilon -4\Lambda $, changes as the result of diminishing of the energy density of matter $\varepsilon $ in the course of the evolution (expansion) of the early cold Universe. As the result, on the inflationary stage of expansion $R\left(t\right)$ reaches a negative critical value, where the first-order phase transition occurs with the release of the latent heat and rapid (during $10^{-31} $~s) heating of matter to the Planck temperature, which may be considered a ``Big Bang'' with power of about $10^{30} $~GW.

\begin{acknowledgments}
The author would like to express his gratitude to G.M.~Zinoviev, I.M.~Krive, S.M.~Ryabchenko, A.V.~Semenov and V.I.~Pentegov for enlightening discussions and useful suggestions.
\end{acknowledgments}

\bibliography{BigBang}

\end{document}